\documentclass[aps,prl,a4paper,twocolumn,superscriptaddress]{revtex4-1}
\usepackage{ucs}
\usepackage[utf8x]{inputenc}
\usepackage{amssymb,amsmath,bm,bbm}
\usepackage{graphicx}
\usepackage{eucal}
\usepackage{multirow}
\usepackage[colorlinks=true,linkcolor=blue, urlcolor=blue]{hyperref}
\usepackage{soul}

\newcommand{\avst}[1]{\left\langle{#1}\right\rangle_{st}}
\newcommand{\dd}{\mathrm{d}}
\newcommand{\bp}{\mathbf{p}}
\newcommand{\bW}{\mathbf{W}}
\newcommand{\bv}{\mathbf{v}}
\newcommand{\bM}{\mathbf{m}}
\newcommand{\bU}{\mathbf{U}}
\newcommand{\bx}{\mathbf{x}}
\newcommand{\bff}{\mathbf{f}}
\newcommand{\AAA}{\mathcal{A}}
\newcommand{\BB}{\mathcal{B}}
\newcommand{\MM}{\mathcal{M}}
\newcommand{\transposed}{\mathsf{T}}
\newcommand{\mean}[1]{\left\langle{#1}\right\rangle}
\newcommand{\float}[2]{{#1}\times 10^{#2}}

\begin{document}

\title{Internal lipid bilayer friction coefficient from equilibrium canonical simulations}

\author{Othmene Benazieb}
\author{Lisa Berezovska}
\author{Fabrice Thalmann}
\affiliation{Institut Charles Sadron, CNRS and University of Strasbourg, 23 rue du Loess, F-67034 Strasbourg cedex 2, France}
\date{\today}

\begin{abstract}
A fundamental result in the theory of Brownian motion is the Einstein-Sutherland  relation between mobility and diffusion constant. Any classical linear response transport coefficient obeys a similar Einstein-Helfand relation. We show in this work how to derive the interleaflet friction coefficient of lipid bilayer by means of an adequate generalisation of the Einstein relation. Special attention must be paid in practical cases to the constraints on the system center of mass position that must be enforced when coupling the system to thermostat.  
\end{abstract}
\maketitle

In 1905 Einstein~\cite{1905_Einstein} and Sutherland~\cite{1905_Sutherland} obtained a relation $D = k_B T \nu$ between the diffusion coefficient $D$ of a Brownian particle, its mobility coefficient $\nu$ (ratio between average drift velocity and drift force) the absolute temperature $T$ and the Boltzmann constant $k_B$. Similar relations were later established for all the usual transport coefficients (viscosity, thermal conduction,\ldots), the Einstein-Helfand relations~\cite{1960_Helfand}. These expressions provide an alternative way to the Green-Kubo relations for the determination of the transport properties, based on the averaged mean square deviation (MSD) of well chosen dynamical observables. The use of Helfand expressions in molecular dynamics (MD) simulation is however not always practical due to system periodic boundary conditions (PBC)~\cite{2007_Viscardy_Gaspard,2007_Viscardy_Gaspard_2}.

A natural question arises as to determining lipid bilayer friction properties in a similar way, \textit{i.e.} by writing and computing the MSD of a carefully chosen  dynamical observable. Such result would be valuable as an alternative to out-of-equilibrium simulation techniques, which are diversely implemented in the commonly used simulation packages.  

\begin{figure}[t]
\centering
\begin{tabular}{c}
\includegraphics[width=0.48\textwidth]{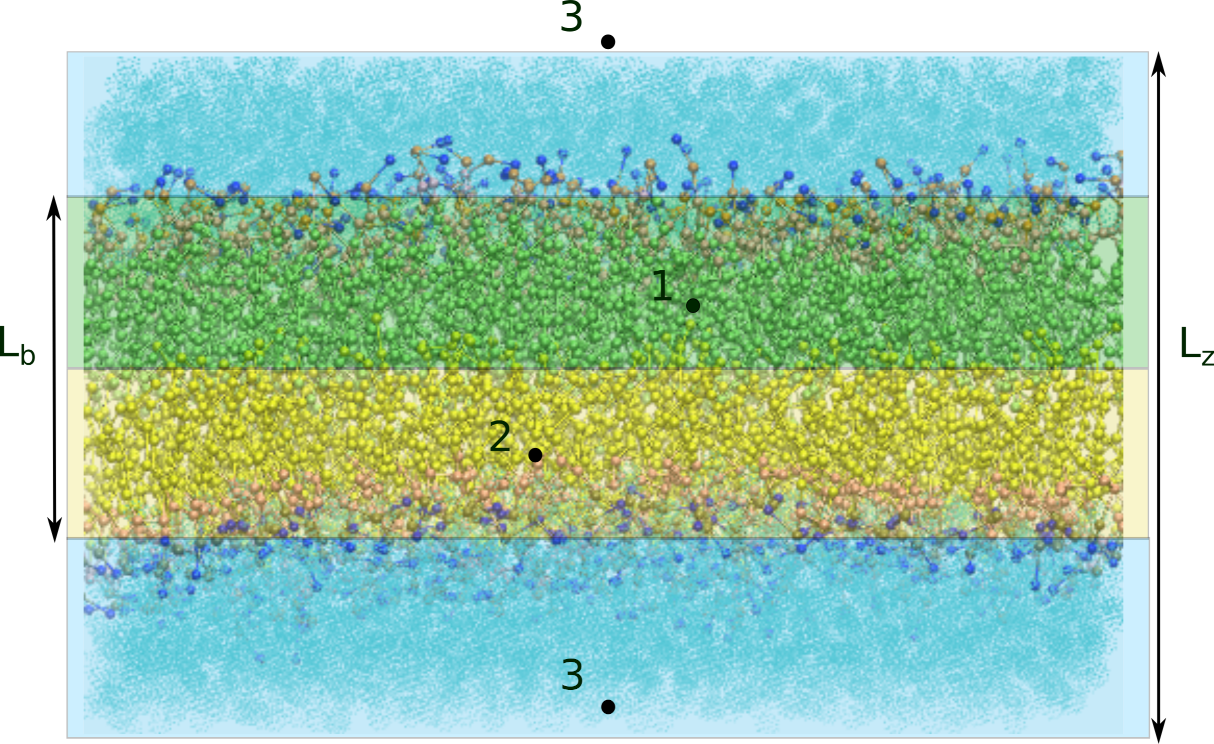}
\end{tabular}
\caption{\label{fig:fig1_slab} 
Coarse-Grained MD snapshot superimposed with the two solid blocks (1,2) and water solvent (3). Each of the 3~components is uniquely located by a single center of mass coordinate along the $x$ direction. The solvent flow is linear-parabolic in the $z$ direction, $L_z$ is the thickness of the (periodic) box, $L_b$ the bilayer thickness, $L_w=L_z-L_b$ the water thickness.}
\end{figure}

A model bilayer system comprising two apposed leaflets and a single water solvent slab is expected  (Fig.~\ref{fig:fig1_slab}) to maintain its self-assembled structure for the longest available simulation times. Two-tails standard lipid molecules are too little soluble in water~\cite{Israelachvili_SurfaceForces,Evans_Wennerstrom_ColloidalDomain} to escape from the bilayer, and have very long leaflet exchange characteristic times~\cite{1971_McConnell_Kornberg,Mouritsen_Bagatolli_MatterOfFat}. Therefore the only molecular motions expected in such a case are the in-plane self diffusion of lipid molecules and bulk diffusion of water molecules. 
Let us then decompose the system into three apposed subsystems: upper lipid leaflet ($\mathcal{S}_1$), lower lipid leaflet ($\mathcal{S}_2$) and solvent ($\mathcal{S}_3$). Denoting $x_i$ the respective horizontal coordinates of the subsystems, one faces the problem of finding a relation between the average displacements covariance matrix $D_{ij}(t)= \langle (x_i(t)-x_j(0)^2\rangle$, $i,j = 1,2,3$, with brackets $\langle\cdot\rangle$ standing for the canonical equilibrium trajectories average, and the desired friction coefficients.

Simulated molecular systems must be coupled to thermostats to generate representative canonical trajectories and keep the system internal energy constant. A number of popular momentum preserving thermostat such as Nose-Hoover chains or V-rescale thermostat requires in turn that the center of mass of the system remains fixed to some arbitrary position, and that no external finite force is applied to the system (mechanical insulation)~\cite{1989_Hoover,2007_Bussi_Parrinello,Frenkel_Smit_MolecularSimulations,Tuckerman_StatisticalPhysics}. In what follows, we adopt this convention throughout. 

The continuous hydrodynamic description of a lipid bilayer system consists in replacing each leaflet by a solid thick slab, and water by a fluid slab at fixed vertical positions (Fig.~\ref{fig:fig1_slab}).  This assumes a low water permeability of the membrane on the one hand (verified in practice) and a system center of mass fixed. We assume $xy$ planar isotropy and restrict ourselves to the $x$ component of the displacements. The hydrodynamic system is characterized by two masses $m_1=m_2=m$ and two velocity scalars $V_1,V_2$ associated to the leaflets, along with a mass density $\rho$ and a continuous velocity field $v_x(z)$ for the water slab.  The water mass $m_w$ and center of mass velocity $V_w$ follow from integrating $\rho$ and $v_x(z)$  along $z$. The solvent flow is assumed to be linear parabolic at all times, a situation covering the Couette and Poiseuille velocity profiles. In the absence of sliding, the flow is completely parametrized by $V_1$,$V_2$ and $V_w$, considered as the slow variables of the many particles system.  When connecting these hydrodynamic variables to molecular simulations, it is necessary to account for possible PBC jumps in the molecular displacements, and to consider continuous, unwrapped trajectories.
Subsystems are possibly acted upon by forces $F_1, F_2,F_3$ in the $x$ direction. Mechanical insulation requires $F_1+F_2+F_3=0$ while the stationary system center of mass imposes $m_1 V_1+m_2 V_2+m_3V_3=0$. As explained in \cite{2021_Benazieb_Thalmann}, the equations of motions of the 3 subsystems read, in the absence of water-lipid bilayer sliding,
\begin{eqnarray}
m_1\dot{V}_1 &=& \left(bA-\frac{2\eta A}{L_w}\right)(V_2-V_1) +\frac{6\eta A}{L_w}(V_3-V_1)+F_1;\nonumber\\
m_2\dot{V}_2 &=& \left(bA-\frac{2\eta A}{L_w}\right)(V_1-V_2) +\frac{6\eta A}{L_w}(V_3-V_2)+F_2;\nonumber\\
m_3\dot{V}_3 &=& \frac{6\eta A}{L_w}(V_1+V_2-2V_3)+F_3,
\label{eq:HydrodynamicForces}
\end{eqnarray}
where $A$ is the area of the slab, $\eta$ the Newtonian viscosity of the solvent and $b$ the interleaflet friction coefficient. The relation expresses that internal forces between components are proportional to the system area, and linearly dependent on the mutual velocity differences. 

The purpose of the Letter is to establish an Einstein-Helfand relation for the frictions coefficients $b$ and $\eta/L_w$. For this purpose we generalize the Langevin-Smoluchovski (overdamped) stochastic equation of motion $\dd(mv) = -\zeta v \dd t + \sqrt{2m\zeta k_BT} \dd W(t)$ of a Brownian particle ($m$ mass, $v$ velocity, $\zeta$ "Stokes"-friction coefficient, $t$ time, $\dd W$ differential of a normalized Wiener process)~\cite{Gardiner_HandbookStochastic}. It is well known in this situation that the Brownian diffusion coefficient equals $k_B T/\zeta$ and that the average kinetic energy $m v^2/2= k_B T/2$ is given by the equipartition theorem. The generalization in the case of the 3 slabs reads:
\begin{eqnarray}
\dd \bp = -\AAA \bp\dd t + \BB \dd \bW(t),
\label{eq:GeneralizedIto}
\end{eqnarray}
$\bp =(m V_1, m V_2, m_w V_3)$ being a vector of impulsions and $\bW(t)$  3~independent normalized Wiener processes. The mass $\MM$ and friction $\AAA$ matrices follow from eq.~\ref{eq:HydrodynamicForces}.
\begin{eqnarray}
\MM &=& \left( 
\begin{array}{ccc}
m & 0 & 0\\
0 & m & 0\\
0 & 0 & m_w
\end{array}
\right) = \MM^{\transposed};\\
\AAA\MM &=& A \left(
\begin{array}{ccc}
 b+ 4\eta/L_w & -b +2\eta/L_w & -6\eta/L_w\\
-b+ 2\eta/L_w & b+4\eta/L_w & -6\eta/L_w\\
-6\eta/L_w & -6\eta/L_w & 12\eta/L_w
\end{array}
\right)\nonumber\\
&=& \MM \AAA^{\transposed}.
\end{eqnarray}
The mass matrix relates the velocities vector $\bv=(V_1,V_2,V_3)$ to the impulsions $\bp = \MM \bv$. The random noise matrix $\BB$ (defined up to a $\mathcal{O}(3)$ orthogonal transformation) must be chosen so that the correct canonical average $\langle \bp \bp^{\transposed}\rangle$ is recovered, ${\transposed}$ representing the real transpose of vectors and matrices. This implies~\cite{Gardiner_HandbookStochastic}
\begin{equation}
\AAA \cdot \langle\bp \bp^{\transposed}\rangle + \langle \bp \bp^{\transposed}\rangle \cdot \AAA^{\transposed} = \BB\BB^{\transposed}.
\label{eq:FDT}
\end{equation}
However, due to the use of a thermostat, a momentum conservation constraint $\bU^{\transposed}\bp =0$ holds, with $\bU=(1,1,1)$. The friction matrix $\AAA$ is only of rank 2 as $\bU^{\transposed}\AAA=0$ (so does $\AAA \MM \bU=0$).  

The total vanishing impulsion constraint modifies the energy equipartition theorem, as the thermal energy $k_BT$ of two degrees of freedom is shared by the three subsystems according to their respective inverse masses. A calculation (Supplemental Material, SM) gives 
\begin{equation}
    \langle \bv \bv^{\transposed}\rangle 
    = k_BT\, \bigg[\MM^{-1}-\frac{\bU\bU^{\transposed}}{m_t}\bigg],
    \label{eq:velocityCovariance}
\end{equation}
with $m_t = m_1+m_2+m_3$ the total mass. The impulsion covariance matrix follows immediately, given $\MM^{\transposed}=\MM$:
\begin{equation}
\langle \bp\bp^{\transposed}\rangle = \MM^{\transposed} \langle \bv \bv^{\transposed}\rangle \MM\nonumber\\
 =k_BT\left[\MM -\MM \frac{(\bU\bU^{\transposed})}{m_t}\MM \right].  
\label{eq:EquipartitionGeneralized}
\end{equation}
Combining eqs.~\ref{eq:FDT} and \ref{eq:EquipartitionGeneralized} along with the constraints $\AAA\MM\bU = 0$ and $\bU^{\transposed}\MM\AAA^{\transposed}=0$ leads to a simple expression for the random force correlations 
\begin{equation}
    \BB\BB^{\transposed} = k_BT(\AAA\MM + \MM\AAA^{\transposed}) = 2k_BT \AAA\MM.
\end{equation}
Meanwhile, the long times (damped) displacement covariance matrix can be obtained by integrating eq.~\ref{eq:GeneralizedIto}, leading to a displacement vector $\Delta \bx=(x_i(t)-x_i(0))$ in terms of the vector of Wiener processes $(W_i(t))$, given by $\AAA\MM\Delta\bx(t) = \BB\bW(t)$ and thus  $\AAA\MM \Delta\bx\Delta\bx^{\transposed}(\AAA\MM)^{\transposed} = \BB \bW\bW^{\transposed}\BB^{\transposed}$. Taking the thermal average leads to 
\begin{equation}
    \AAA\MM \langle \Delta\bx\Delta\bx^{\transposed}\rangle \AAA \MM = \BB\BB^{\transposed} t = 2\,k_BT\, \AAA\MM\, t.
    \label{eq:DisplacementCovarianceResult}
\end{equation}
Eq.~\ref{eq:DisplacementCovarianceResult} formally solves the problem, by connecting the covariance displacement matrix on the left hand side with the friction matrix $\AAA\MM$ on the right hand side. It represents the desired Einstein-Helfand expression for the lipid bilayer frictions. Nevertheless, the expression is not useful as such, due to $\AAA\MM$ being a rank~2 matrix. It cannot be explicitly inverted to yield the desired displacement covariance matrix alone on the left hand side of an equation. Eq.~\ref{eq:DisplacementCovarianceResult} takes actually a Moore-Penrose pseudo-inverse matrix form. 

In the case of interest, it is possible to express the covariance matrix using the orthogonal change of basis 
\begin{widetext}
\begin{equation}
    O = \frac{1}{\sqrt{3}}\left(
    \begin{array}{ccc}
     \frac{1+\sqrt{3}}{2} &
     \frac{1-\sqrt{3}}{2} & 1\\
     \frac{1-\sqrt{3}}{2} & \frac{1+\sqrt{3}}{2} & 1\\
     -1 & -1 & 1
    \end{array}
    \right)
\; ; \; 
    O^{\transposed} \AAA\MM O =A \left(
    \begin{array}{ccc}
    b+\frac{10\eta}{L_w} & \frac{8\eta}{L_w} -b& 0\\
     \frac{8\eta}{L_w} -b & b+\frac{10\eta}{L_w} & 0 \\
    0 & 0 & 0
    \end{array}
    \right).
\end{equation}
Parameterizing the displacement covariance matrix $\mathcal{D}$ with Voigt indices 
4,5,6, $D_1=D_2$ and $D_4=D_5$ 
\begin{equation}
    \left(
    \begin{array}{ccc}
    \langle (\Delta \bx_1)^2 \rangle & 
    \langle \Delta\bx_1\Delta\bx_2 \rangle &
    \langle \Delta\bx_1\Delta\bx_3 \rangle \\
    \langle \Delta\bx_2\Delta\bx_1\rangle  & 
    \langle (\Delta\bx_2)^2\rangle &
    \langle \Delta\bx_2\Delta\bx_3\rangle \\
    \langle \Delta\bx_3\Delta\bx_1\rangle &
    \langle \Delta\bx_2\Delta\bx_3\rangle &
    \langle (\Delta\bx_3)^2 \rangle 
     \end{array}
    \right) 
    =
    2\left(
    \begin{array}{ccc}
    D_1 & D_6 & D_4\\
    D_6 & D_1 & D_4 \\
    D_4 & D_4 & D_3
    \end{array}
    \right) t,
\end{equation}
leads to the explicit inverse relation
\begin{equation}
    \frac{3k_BT}{A} \left(
    \begin{array}{cc}
    1 & 0 \\ 0 & 1
    \end{array}
    \right)
    = 
    \left(
    \begin{array}{cc}
    b+\frac{10\eta}{L_w} & \frac{8\eta}{L_w}-b \\ 
    \frac{8\eta}{L_w}-b & b+\frac{10\eta}{L_w}
    \end{array}
    \right)\cdot
    \left(
    \begin{array}{cc}
    (2D_1 -D_6 +D_3-2D_4)\, &  (2D_6-D_1+D_3-2D_4) \\ 
     (2D_6-D_1+D_3-2D_4)\, & (2D_1 -D_6 +D_3-2D_4)
    \end{array}
    \right).
\label{eq:ExplicitHelfandExpression}
\end{equation}
%
\end{widetext}
Equation~\ref{eq:ExplicitHelfandExpression} is a $2\times 2$ matrix generalization of the Stokes-Einstein relation, and constitutes the main result of this Letter. It is easy to establish two further relations between the covariance parameters 
\begin{equation}
m_w D_3 + 2m D_4 = 0\; ; \; m (D_1+D_6) + m_w D_4=0,  
\label{eq:RelationDCoefficients}
\end{equation}
showing that only two independent degrees of freedom are left in the displacement covariance matrix to match the two independent degrees of freedom of the friction matrix $b$ and $\eta/L_w$. Eq.~\ref{eq:RelationDCoefficients} follows from the time integration of $\bv\MM\bU=0$, proving that the displacement vector $\Delta\bx(t)$ stays always orthogonal to $\MM\bU$. Note that the presentation of eq.~\ref{eq:ExplicitHelfandExpression} is not unique, due to a O(1) degeneracy associated with the choice of the orthogonal matrix $O$, whose sole effect is to rotate the 2 coordinates in \ref{eq:ExplicitHelfandExpression}.

We now check that the Brownian description proposed predicts correctly the behavior of a simulated lipid membrane system (coarse-grained Martini model~\cite{2007_Marrink_deVries}, 512 lipids, 10~$\mu$s simulated, see SM for details). Predictions for the velocity covariance matrix can be assessed by estimating the deviation $\varepsilon_{\mathrm{ept}}$  from the expected result, using the matrix norm $||X||_q^2 = \mathrm{tr}(XX^{\transposed})$ (see eq~\ref{eq:EquipartitionGeneralized}):
\begin{equation} 
\varepsilon_{\mathrm{ept}}=\left\lVert k_BT\left[\MM^{-1}-\frac{\bU\bU^{\transposed}}{m_t}\right] -\langle \bv \bv^{\transposed}\rangle\right\rVert_q / \left\lVert \langle \bv \bv^{\transposed} \rangle\right\rVert_q.
\end{equation}
We find good agreement with the prediction as $\varepsilon_{\mathrm{ept}}\simeq  \float{1.8}{-3}$ is of the order of two parts per thousand. Using the same data and the  Einstein-Helfand relation (\ref{eq:ExplicitHelfandExpression}) we obtained respectively $b=\float{2.54}{6}~\mathrm{Pa\,s\,m}^{-1}$ and $\eta = \float{8.1}{-4}~\mathrm{Pa\,s}$, with an estimated relative accuracy  of the order of 0.05 (see SM). This compares favorably with the reported values, for the same system and conditions, of $b=\float{2.55\pm 0.10}{6}~\mathrm{Pa\,s\,m}^{-1}$ and $\eta = \float{8.}{-4}~\mathrm{Pa\,s}$~\cite{2021_Benazieb_Thalmann}. 

We have so far shown how the standard Wiener process for damped Brownian motion generalizes to the case of constrained vanishing total momentum. In doing so, one finds eq.~\ref{eq:DisplacementCovarianceResult} as the generalization of the 2$^{nd}$ fluctuation-dissipation theorem. A natural question arises as whether it is possible to write an equivalent version of the 1$^{st}$ fluctuation-dissipation  theorem (FDT), which relates the linear response of a system to an equilibrium correlation function. The unbounded free Brownian motion is unfortunately not an equilibrium situation, with the 1st FDT violated~\cite{1994_Cugliandolo_Parisi}. It is therefore necessary to place the system under conditions consistent with a stationary thermal equilibrium state. The harmonic confinement potential is the simplest and most natural example. Once the FDT proven in the harmonic case, it can in principle be extended perturbatively to nonlinear analytic potentials~\cite{1975_Deker_Haake,1996_Bouchaud_Mezard}.

\begin{figure}[b]
\centering
\includegraphics[width=0.36\textwidth]{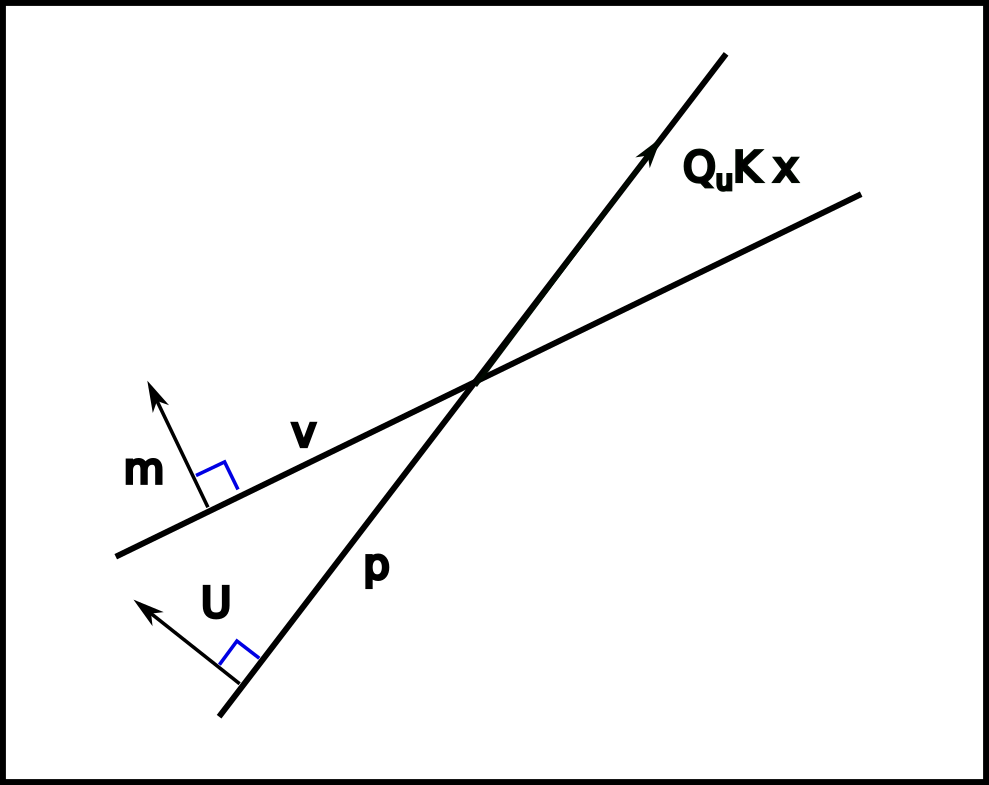}
\caption{\label{fig:fig2_projectionLinearForce} Illustration of the projection procedure. Velocities are constrained to an hyperplane $\mathbf{m}^{\transposed}\bv=0$, momenta are constrained to the hyperplane $\bU^{\transposed}\bp=0$. To satisfy the conservation of the center of mass position, the linear force $\mathcal{K}\bx$ must be projected onto the momenta hyperplane. In simulations, the constraint on the system center of mass position acts by removing the out-of-plane component of the force. }
\end{figure}

For this purpose, we now consider the damped Brownian (Smoluchovski) process (\ref{eq:GeneralizedIto}) in the presence of a linear force field 
\begin{equation}
\mathcal{A}\mathcal{M}\dd \bx +\mathcal{K} \bx \dd t= \BB \dd \bW(t)+\bff(t)\dd t
\label{eq:brownianConfined}
\end{equation}
with $\bx =(x_1(t),x_2(t),x_3(t))$ the stochastic trajectory, $\bp = \mathcal{M}\bx$, $\mathcal{K} = \mathcal{K}_{ij}$ a symmetric curvature matrix  representing the linear conservative force-field and $\bff(t)$ the arbitrary time dependent external force conjugated to the average position response $\mean{\bx(t')}$ at posterior times $t'>t$. The harmonic and perturbation forces must both be consistent with a total vanishing (conserved total momentum) constraint $\bU^{\transposed}\mathcal{K} = \mathcal{K}\bU =0$ and $\bU^{\transposed}\bff(t)=0$. The motion is restricted to the orthogonal subspace $U^{\transposed}\mathcal{M}\bx=0$, denoted $m^{\perp}$, while all forces and momenta belongs the orthogonal subspace $U^{\perp}$. Positiveness of the harmonic potential requires $\bx^{\transposed}\mathcal{K}\bx >0$. The problem reduces to computing the correlation and response properties of the projected vector $\mathcal{Q}_u\bx(t)$ orthogonal to $\bU$, with $\mathcal{Q}_u = \mathbbm{1}-\bU\bU^{\transposed}/3$ (Figure~\ref{fig:fig2_projectionLinearForce}). Expressing the original displacement $\bx$ in terms of the projected $\mathcal{Q}_u\bx(t)$ is straightforward:
\begin{equation}
    \bx = \left( \mathbbm{1}-\frac{\bU\bM^{\transposed}}{m_t}\right)\mathcal{Q}_u\bx.
    \label{eq:retroProjection}
\end{equation}
Let us now determine the statistical properties of $\mathcal{Q}_u\bx$. We first observe that  $\mathcal{Q}_u\mathcal{A}\mathcal{M}\mathcal{Q}_u = \mathcal{A}\mathcal{M}$ and $\mathcal{K}\mathcal{Q}_u = \mathcal{K}$. Eq.~\ref{eq:brownianConfined} can be rewritten
as
\begin{equation}
    \mathcal{Q}_u\mathcal{A}\mathcal{M}\mathcal{Q}_u \mathcal{Q}_u \dd\bx +\mathcal{K}\mathcal{Q}_u \bx\dd t = \mathcal{B} \dd\bW(t) +\bff(t) \dd t.
    \label{eq:projectedStochastic}
\end{equation}
The projected dissipation matrix $\mathcal{Q}_u\mathcal{A}\mathcal{M}\mathcal{Q}_u$ is invertible in the vector subspace $U^{\perp}$ orthogonal to $\bU$, and its rank~2 inverse denoted $\mathcal{J}_u = \{\mathcal{Q}_u\mathcal{A}\mathcal{M}\mathcal{Q}_u\}^{-1}$. All terms in~\ref{eq:projectedStochastic} belong to $U^{\perp}$. The projected stochastic solution reads
\begin{eqnarray}
    \mathcal{Q}_u\bx(t) &=& \exp(-\mathcal{J}_u\mathcal{K} t)\mathcal{Q}_u\bx(0) +\\
    & & +\int_0^{t} \exp(-\mathcal{J}_u\mathcal{K} (t-s)) \mathcal{J}_u
[\mathcal{B}\dd \bW(s) + \bff(s)\dd s],\nonumber
\end{eqnarray}
from which the undriven $\bff=0$ correlation for $t>0$
\begin{eqnarray}
\mathcal{C}_u(t) &= &\mean{\mathcal{Q}_u\bx(t)\bx(0)^{\transposed}\mathcal{Q}_u}\nonumber\\
&=& \exp(-\mathcal{J}_u\mathcal{K}t) \mean{\mathcal{Q}_u\bx(0)\bx(0)^{\transposed}\mathcal{Q}_u} 
\end{eqnarray}
and the causal response $\mathcal{R}_u(t)=\exp(-\mathcal{J}_u\mathcal{K} t) \mathcal{J}_u$, defined by $\delta \mean{\mathcal{Q}_u\bx(t)} = \int_0^t \mathcal{R}_u(t-s) \delta\bff(s) \dd s$ can be obtained.
Similar reasoning as eq.~\ref{eq:DisplacementCovarianceResult} leads to a stationary covariance matrix
\begin{equation} 
\lim_{t\to\infty} \mean{\mathcal{Q}_u\bx(t)\bx(t)^{\transposed}\mathcal{Q}_u}= \avst{\mathcal{Q}_u\bx \bx^{\transposed}\mathcal{Q}_y} = k_BT \{\mathcal{Q}_u\mathcal{K}\mathcal{Q}_u\}^{-1}, 
\end{equation}
ensuring energy equipartition into the two harmonic potential degrees of freedom. 
The stationary projected correlation function therefore simplifies as 
\begin{widetext}
\begin{equation}
    \mathcal{C}_u(t) = \mean{\mathcal{Q}_u\bx(t)\bx(0)^{\transposed}\mathcal{Q}_u}
    = k_B T \left\{ H(t)\,\exp(-\mathcal{J}_u\mathcal{K} t) \{\mathcal{Q}_u\mathcal{K}\mathcal{Q}_u\}^{-1} + H(-t)\, \{\mathcal{Q}_u\mathcal{K}\mathcal{Q}_u\}^{-1} \exp(\mathcal{K}\mathcal{J}_u t)\right\}
\end{equation}
%
\end{widetext}
where a Heaviside function $H(t)$ distinguishes positive and negative time values, and account for a possible non commutation of the $\mathcal{A}\mathcal{M}$ and $\mathcal{K}$ matrices. 
It is then clear that, in this form, a first fluctuation dissipation theorem 
\begin{equation}
    \frac{\dd\mathcal{C}_u}{\dd t} = k_B T \left\{ H(t) \mathcal{R}_u(t) - H(-t)\mathcal{R}_u^{\transposed}(-t) \right\} 
\end{equation}
holds. Expressing the relation between $\mathcal{C}(t)=\langle\bx(t)\bx(0)^{\transposed}\rangle$ and $\mathcal{R}(t) = \delta\mean{\bx(t)}/\delta \bff(0)$ using \ref{eq:retroProjection} is immediate. 

To conclude, we have introduced an original equilibrium fluctuation relation between the center of mass mutual diffusion coefficients of a simulated lipid membrane with periodic boundary conditions and the viscous dissipation coefficients (interleaflet friction and solvent viscosity) relevant to the motion in the bilayer plane. This result assumes that no lipid exchange takes place, and that solvent penetration into the membrane can be neglected. It is consistent with the use of a thermostat where the center of mass of the whole system is forced to be static. It is based on a macroscopic long range and long times hydrodynamic description of the mutual bilayer components displacements. It also disregards any sliding of the solvent at the bilayer interface, usually considered as negligible as a first approximation. The diffusion parameters introduced in the discussion depends on the area of the simulated system, very much like the Stokes sphere mobility depends on its radius. A compromise must be found between increasing the system size to reach some hydrodynamic limit, and keeping it small enough to prevent Helfrich undulations~\cite{Safran_Surfaces} and preserving enough leaflet Brownian diffusion.
With very little extra-cost in terms of simulation and analysis, this result is poised to become a standard characterization of realistic numerical membranes, provided they are simulated long enough for the leaflet center of mass displacements to be estimated.\\
\textit{Acknowledgement} The authors would like to acknowledge the High Performance Computing Center of the University of Strasbourg for supporting this work by providing scientific support and access to computing resources (grant g2021a337c). 



\end{document}


\title{Internal lipid bilayer friction coefficient from equilibrium canonical simulations.\\Supplemental Material}

\author{Othmene Benazieb, Lisa Berezovska, Fabrice Thalmann\\
Institut Charles Sadron, CNRS and University of Strasbourg,\\
23 rue du Loess, F-67034 Strasbourg cedex 2, France}
\date{\today}

\maketitle

\subsection*{Simulation conditions}

Our simulations are based on the Martini~v2.0 lipid coarse-grained force field~\cite{2007_Marrink_deVries}, and use 512 distearoylphosphatidylcholine (DSPC, 5 beads chain model) along with 5120 water bead molecules. The DSPC bilayer is numerically fluid at 340~K, simulated temperature. 

The MD simulation engine was Gromacs~2018.4~\cite{2015_Abraham_Lindahl}. After an initial equilibration stage, we let the system evolve in contact with a V-rescale thermostat (coupling time 1~ps) with separately coupled DSPC and water groups and a Parinello-Rahman semiisotropic barostat (coupling time 12~ps, compressibility $\dec{3}{-4}$~bar$^{-1}$). The integration time step was $\dd t=10$~fs. A total of 25 consecutive trajectories of 400~ns each, cumulating to 10~$\mu$s were used in the statistics presented here. The 25 trajectories were used combined into a bootstrap statistical analysis of the variability of the presented results.

Trajectory file in Gromacs binary format \texttt{trr} were created recording center of forces positions and velocities every 20~ps. Home made~C softwares based on  Gromacs libraries were used to manipulate index and trajectory files, taking care of periodic boundary conditions jumps, computing observables such as leaflet and solvent center of mass positions and determining instantaneous kinetic energies. Mean squared displacements were computed from the subsystem center of mass trajectories.

\subsection*{Equipartition of energy}

The system comprises three spatially separated components: the upper and lower leaflets and the water slab. Molecular exchange between components, for a phospholipid bilayer simulated  under normal conditions of pressure and temperature, is marginal, corresponding mostly to water beads transits across bilayers. An instantaneous center of mass velocity can be defined. The canonical equilibrium distribution is supposed to be valid, which can be can be explicitly verified in our case, see eq~\ref{eq:numericalVelocities}.

For the sake of generality, we address first the situation with three arbitrary masses $\mathbf{m}=(m_1,m_2,m_3)$. In the Maxwell velocity distribution, the three euclidean directions are uncoupled, and it is sufficient to restrict ourselves the $x$ component. This amounts to considering a center of mass velocity vector $\bv=(V_1,V_2,V_3)$, subject to a kinematic constraint $\bM^{\transposed}\bv=0$. 

The instantaneous center of mass velocity covariance matrix 
\begin{equation}
    \mean{v_iv_j} = \frac{ \int \dd\bv\, (v_iv_j)\delta( \mathbf{m}^{\transposed}\bv) \exp[-\frac{\beta}{2}(m_1v_1^2+m_2v_2^2+m_3v_3^2)]}
    {\int \dd\bv\, \delta( \mathbf{m}^{\transposed}\bv) \exp[-\frac{\beta}{2}(m_1v_1^2+m_2v_2^2+m_3v_3^2)]}
\end{equation}
can be computed in a number of different manners, all leading to the same result. It turns out that the computation of the impulsions covariance matrix is simpler. 
\begin{equation}
    \mean{p_ip_j} = \frac{ \int \dd\bp\, (p_ip_j)\delta( \bU^{\transposed}\bp) \exp[-\frac{\beta}{2}(\frac{p_1^2}{m_1}+\frac{p_2^2}{m_2}+\frac{p_3^2}{m_3})]}
    {\int \dd\bp\, \delta( \bU^{\transposed}\bp) \exp[-\frac{\beta}{2}(\frac{p_1^2}{m_1}+\frac{p_2^2}{m_2}+\frac{p_3^2}{m_3})]}
\end{equation}
One takes advantage of the mass matrix $\MM$ to express the kinetic energy $\bp^{\transposed}\MM^{-1}\bp/2$ and one uses the following  gaussian regularization of the $\delta$:
\begin{equation}
\delta(\bU^{\transposed}\bp) = \lim_{\varepsilon \downarrow 0}\frac{1}{\sqrt{2\pi\varepsilon}}\, e^{-\displaystyle\frac{\bp^{\transposed}\bU\bU^{\transposed}\bp}{2\varepsilon}}
\end{equation}
A classical result on multidimensional gaussian integrals states that 
\begin{eqnarray}
    \mean{\bp \bp^{\transposed}} &=& \lim_{\varepsilon \downarrow 0} \left(\beta \MM^{-1} +\frac{\bU\bU^{\transposed}}{\varepsilon}\right)^{-1} \nonumber\\
&=& \lim_{\varepsilon \downarrow 0}  \varepsilon \left(\varepsilon\beta \MM^{-1} +\bU\bU^{\transposed}\right)^{-1}
\end{eqnarray}
Guessing for a matrix inverse form $C_1\,\MM + C_2\,\bM\bM^{\transposed}$ with  $C_1,C_2$ numerical constants to determine leads to 
\begin{equation}
    (\varepsilon\beta \MM^{-1} +\bU\bU^{\transposed})(C_1\,\MM + C_2\,\bM\bM^{\transposed})= 
    (\varepsilon\beta C_1)\mathbbm{1}+ (C_1+m_t C_2+\varepsilon\beta C_2)\bU\bM^{\transposed}
\end{equation}
with $m_t=\mathrm{tr}(\MM)$ representing the total mass. This implies $\varepsilon\beta C_1=1$ and $C_1+(m_t+\varepsilon\beta)C_2=0$. The matrix inverse is therefore equal to 
\begin{equation}
\frac{1}{\beta\varepsilon}\left(\MM -\frac{\bM\bM^{\transposed}}{m_t+\varepsilon\beta}\right)
\end{equation}
Then finally in the $\varepsilon\to 0$ limit
\begin{eqnarray}
    \mean{p_i p_j} &=& k_B T \left(\MM -\frac{\bM\bM^{\transposed}}{m_t}\right)\\
    \mean{v_i v_j} &=& k_B T \left(\MM^{-1} -\frac{\bU\bU^{\transposed}}{m_t}\right)
\end{eqnarray}
This result is valid for any number of mass components, provided the system is only subject to one kinetic constraint $\bU^{\transposed}\bp=\bM^{\transposed}\bv=0$. 

\subsection*{Numerical verification of the equipartition of energy}

In the case of the simulated Martini lipid bilayer, the numerical velocity covariance matrix in nm$^2$ps$^{-2}$ reads,  with 2\textperthousand\, relative confidence
\begin{equation}
\mean{\bv \bv^{\transposed}}=\left(
    \begin{array}{ccc}
    \dec{7.740}{-6} & \dec{-3.190}{-6} & \dec{-3.184}{-6}\\
    \dec{-3.190}{-6} & \dec{7.747}{-6} & \dec{-3.189}{-6}\\
    \dec{-3.184}{-6} & \dec{-3.189}{-6} & \dec{4.461}{-6}
    \end{array}
    \right)
    \label{eq:numericalVelocities}
\end{equation}
This compares favorably with the theoretical prediction, using $T=340$K, $m_1=m_2=256\times 14 \times 72=258048$~g mol$^{-1}$, $m_3=m_w = 5140\times 72 = 368640$~g mol$^{-1}$ and $m_t = 884736$~g mol$^{-1}$, expressed in the same unit as above.
\begin{equation}
k_BT\, \bigg[\MM^{-1}-\frac{\bU\bU^{\transposed}}{m_t}\bigg] =
\left(
    \begin{array}{ccc}
    \dec{7.755}{-6} & \dec{-3.193}{-6} & \dec{-3.193}{-6}\\
    \dec{-3.192}{-6} & \dec{7.755}{-6} & \dec{-3.193}{-6}\\
    \dec{-3.193}{-6} & \dec{-3.193}{-6} & \dec{4.470}{-6}
    \end{array}
    \right)
\end{equation}

\subsection*{Diffusive behavior of the center of mass}

\begin{figure}[t]
    \centering
    \resizebox{0.65\textwidth}{!}{\includegraphics{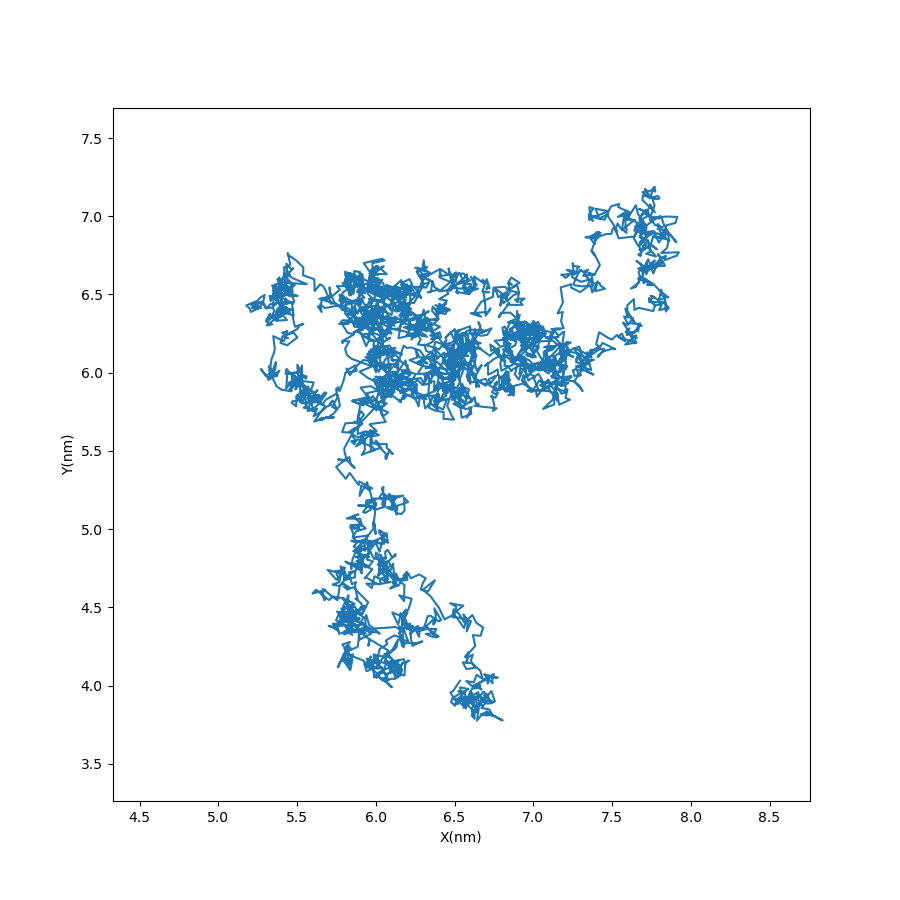}}
    \caption{Horizontal projection of the Brownian diffusion of the center of mass of the upper leaflet, every 10~ps, during 400~ns.}
    \label{fig:brownTrajDown}
\end{figure}

\begin{figure}[t]
    \centering
    \resizebox{0.8\textwidth}{!}{\includegraphics{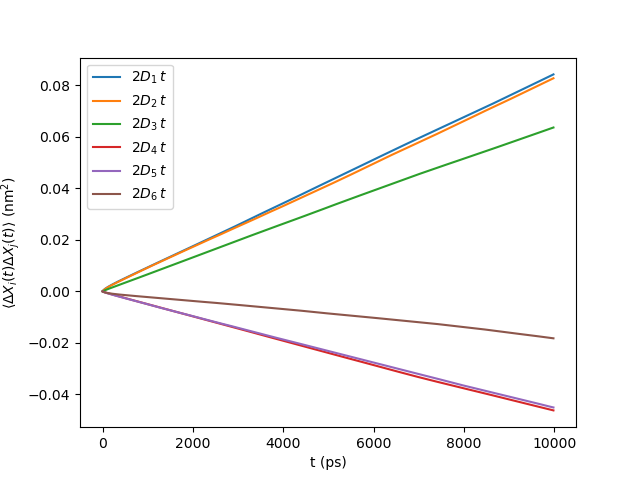}}
    \caption{Average displacements correlation matrix $\mean{\Delta X_i(t)\Delta X_j(t)}$, averaged over 10~$\mu$s. Diffusion coefficients $D_1,\ldots D_6$ can be estimated from a linear fit of the second half of the curves (100-200~ps).}
    \label{fig:mutualDiffusion}
\end{figure}


Figure~\ref{fig:brownTrajDown} represents a sample of Brownian center of mass equilibrium displacement of one leaflet in the $x,y$ plane.  Parameterizing the displacement covariance matrix $\mathcal{D}$ as 
\begin{equation}
    \left(
    \begin{array}{ccc}
    \langle (\Delta \bx_1)^2 \rangle & 
    \langle \Delta\bx_1\Delta\bx_2 \rangle &
    \langle \Delta\bx_1\Delta\bx_3 \rangle \\
    \langle \Delta\bx_2\Delta\bx_1\rangle  & 
    \langle (\Delta\bx_2)^2\rangle &
    \langle \Delta\bx_2\Delta\bx_3\rangle \\
    \langle \Delta\bx_3\Delta\bx_1\rangle &
    \langle \Delta\bx_2\Delta\bx_3\rangle &
    \langle (\Delta\bx_3)^2 \rangle 
     \end{array}
    \right) 
    =
    2\left(
    \begin{array}{ccc}
    D_1 & D_6 & D_4\\
    D_6 & D_1 & D_4 \\
    D_4 & D_4 & D_3
    \end{array}
    \right) t,
\end{equation}
Figure~\ref{fig:mutualDiffusion} shows the asymptotic linear diffusion regime of the mutual quadratic displacements along the $x$ direction. In addition, averaging the relations $\Delta X_1(t) (m \Delta X_1(t)+ m\Delta X_2(t)+m_w \Delta X_3(t))=0$ and 
$\Delta X_3(t)(m \Delta X_1(t)+ m\Delta X_2(t)+m_w \Delta X_3(t))=0$ leads to two additional relations
\begin{eqnarray}
m_w D_3 + 2m D_4 &=& 0;\nonumber\\
m (D_1+D_6) + m_w D_4 &=& 0.  
\end{eqnarray}
One typical numerical simulation  gives the following list of values:
\begin{center}
  \begin{tabular}{|c|c|}
    \hline
    \multicolumn{2}{|c|}{Covariant diffusion coefficients ($\mu$m$^2$s$^{-1}\vphantom{\Big\lbrace}$)}\\
    \hline
    $D_1=4.19\pm 0.17$ & $D_4=-2.35\pm 0.13$\\
    \hline
    $D_2=4.11\pm 0.20$ & $D_5=-2.41\pm 0.11$\\
    \hline
    $D_3=3.33\pm 0.10$ & $D_6=-0.75\pm 0.13$\\
    \hline
    \end{tabular}
\end{center}

\begin{eqnarray}
1 + \frac{m}{m_w} \frac{D_4+D_5}{D_3} &\simeq & \dec{-8.4}{-5};\nonumber\\
1 + \frac{m}{m_w}\frac{(D_1+D_2+2 D_6)}{D_4+D_5} & \simeq & \dec{-1.1}{-4},  
\end{eqnarray}
confirming the validity of the approach.

\subsection*{Estimation of the confidence interval}

The bootstrap method consists in randomly resampling a set of randomly fluctuating data in to estimate their intrinsic variability. It is rigorously established  when the fluctuating data are ensured to be statistically independent and gaussian. Otherwise it provides us with a systematic estimate of the sample to sample variability of an average quantity. 

Here we have $N_{r}= 25$ consecutive runs indexed with $\alpha$ at our disposal, which we assume to be essentially weakly correlated (though there are always long lived hydrodynamic fluctuations)~\cite{Press_Flannery_NumericalRecipes_C}. Each run gives a satisfactory estimate of the fitted $D_i^{(\alpha)}$ or friction $b^{(\alpha)},\eta^{(\alpha)}$ coefficients. 50 synthetic samples are build from the reference $\{\alpha\}$ sample set by drawing at random, and with replacement 25 values. The corresponding average values of the synthetic samples provides us with a variance $\sigma^2(\overline{D_i})$, $\sigma^2(\overline{b})$, $\sigma^2(\overline{\eta})$ that we take as a basis for our "2$\sigma$" confidence intervals.

